\documentstyle[12pt]{article}
\makeatletter
 
 \@addtoreset{equation}{section}
\makeatother
\begin{document}
\newcommand{\be}{\begin{equation}}
\newcommand{\ee}{\end{equation}}
\newcommand{\vth}{\vspace{3mm}}
\newcommand{\osp}{osp(2$ \vert $2)$\oplus$osp(2$\vert$2)~}
\newcommand{\Osp}{Osp(2$ \vert $2)$\otimes$Osp(2$ \vert $2)~}
\newcommand{\bec}{\begin{center}}
\newcommand{\ec}{\end{center}}
\newcommand{\befl}{\begin{flushleft}}
\newcommand{\efl}{\end{flushleft}}
\newcommand{\hed}{\quad \quad}
\newcommand{\dfrac}[2]{%
        \frac{\displaystyle{#1}}{\displaystyle{#2}}}
\newcommand{\mapdown}[1]{\Big\downarrow
   \rlap{$\vcenter{\hbox{$\scriptstyle#1\,$}}$ }}
\newcommand{\mapright}[1]{%
  \smash{\mathop{%
   \hbox to 0.5cm{\rightarrowfill}}\limits^{#1}}}
\newcommand{\mapleft}[1]{%
      \smash{\mathop{%
        \hbox to 0.5cm{\leftarrowfill}}\limits^{#1}}}
\begin{flushright}
RUP-4-95\\
April, 1995
\end{flushright}
\vspace*{1.7cm}
\bec
\large{ GEOMETRICAL ASPECT OF \\
             TOPOLOGICALLY TWISTED 2-DIMENSIONAL \\
               CONFORMAL SUPERALGEBRA}\\[1.5cm]
\large{\sc Noriaki Ano}$~ ^\star$
\footnote[0]{$\star$ e-mail: nano@rikkyo.ac.jp}\\[0.3cm]
\normalsize{\it Department of Physics, Rikkyo University,}\\
\normalsize{\it Nishi-Ikebukuro, Tokyo171, Japan}\\
\ec
\normalsize
\vspace{1.0cm}
\bec {\sc ABSTRACT} \ec
We study the topologically twisted \osp
conformal superalgebra.
The algebra includes the Lagrangians
which are intrinsic to the topological field theory and
composed of fermionic generators.
Studying the Lagrangians through
a gauge system of \osp,
geometrical features
inherent to the algebra are revealed:
a moduli space associated with the algebra is derived
and the crucial roles which the fermionic generators play in
the moduli space are clarified.
It is argued that there exists a specific relation
between the topological twist and the moduli problem
through a geometrical aspect of the algebra.
\newpage
\section{\rm Introduction}
\hed
In the recent progress of the quantum field theory (QFT),
the detection of the cohomological field theory
may be most fascinating development.
The theory is a kind of the topological QFT
which deal with topological invariants, and
has been pioneered by E. Witten\cite{e.witten1}.
We refer to the cohomological field theory
as TFT in the present paper
and will focus on it.
The theory has some characteristic properties on the construction
and has distinct framework.
Therefore, many energetic researches in TFT have been
done\cite{d.birmingham&m.blau&m.rakowski&g.thompson1}
and then TFT
has been proved to be a real solid methodology in QFT.
A few substantial problems associated with TFT still remain to be solved,
for example, about the topological twist, however.
In the conformal field theory (CFT),
the toplogical twist of N=2 CFT is performed through a redefinition
of the energy-momentum tensor of N=2 theory\cite{t.eguchi&s.k.yang1},
which generates the bosonic CFT models of vanishing
central charge with hidden fermionic (topological) symmetry.
In relation to the topological twisting mechanism,
the different twistings of the same model yield
the different moduli problems, respectively,
which are related through the mirror symmetry\cite{mirror}
as the explicit example of twisting in general.
For another example,
the topological gauged WZW models\cite{h-l.hu1} are composed of two
different gauge fixing procedures from
the same bosonic model, not necessarily
twisting of the N=2 supersymmetry\cite{e.witten2}
of Kazama-Suzuki model\cite{y.kazama&h.suzuki1},
and in this case there surely exists
the mirror symmetry.

There are two typical stand points
for constructing TFT, i.e.
topological twisting and BRST gauge fixing.
Both approaches result in the so-called
moduli problem\cite{e.witten3}\cite{j.sonnenschein1}.
In either case, the remarkable characteristic is that the Lagrangian
is described as ${\cal L}=\{ {\cal Q},\enskip \star\}$,
where ${\cal Q}$ is the fermionic
operator of nilpotency, i.e. the so-called topological symmetry.
In terms of the ordinary
QFT words, ${\cal L}$ is just composed of the BRST gauge
fixing and the FP ghost terms, and the ${\cal Q}$
corresponds to the BRST operator.
Because of the BRST-exact form of ${\cal L}$,
every correlation function is independent
of the coupling factor
as a consequence of which the leading contribution to the path-integral
is only the classical configuration of the fields, i.e. zero mode.
This zero mode configuration is associated with some moduli space.
In the BRST approach, the relation between the moduli problem
and TFT may be comparatively
clear owing to the intrinsic constructing
procedure where some moduli problem can be
settled as the gauge fixing condition.
In the topological twisting formalism,
the above relation is not much clear, on the contrary.
It seems that there has not been a
common recognition on what the topological twist is really doing.

In the present paper,
we concentrate on N=2 finite-dimensional superalgebra
in two dimensions, perform the topological
twist on such a superalgebra,
and show a characteristic property
associated with the topological twist
by discussing the twisted superalgebra,
i.e. the so-called topological algebra, through a gauge system.
First, a geometrical feature
inherent to the algebra is revealed,
and then it is argued that there exists a specific relation
between the topological twist and the moduli problem
through a geometrical aspect of the algebra.

In the next section, we decide on \osp as N=2 finite-dimensional
superalgebra and perform the topological
twist on \osp, so that topological algebra is obtained.
In section 3,  three types of the TFT's Lagrangian are found
in the topological algebra.
One of the three types of Lagrangian is focused on
and the field configuration is investigated in the case
of zero-limit of the coupling factor on the path-integral
by considering a gauge system.
It is shown that
this configuration is indeed a moduli space of flat connections
associated with the topological algebra,
and this fact originates from vanishing Noether current.
In section 4, a geometrical aspect of the fermionic
charges is discussed.
Under the weak coupling limit,
the total Lagrangian which is a linear
combination of the three Lagrangians
is regarded as the Laplacian operator on the moduli space,
and the fermionic charges as Fredholm operators.
Taking account of these facts, the moduli space which is
obtained formally in sect.3
is made more visual.
It is also shown that the index of these operators
could be derived if proper support in the
moduli space can be defined.
Lastly in sect.4, we discuss the triviality of the path-integral
and obtain a non-trivial TFT's observable.
The fact will supports the argument developed
in the present paper.
In the final section,
It is claimed that the algebra has
a specific relation
with the moduli problem
and the some remark about the vanishing Noether current
which plays a crucial role in
the following discussions is mentioned.\\
\section{\rm Procedure of Topological Twist}
\befl \underline{ 2-1 \ \osp Algebra} \efl
\hed The first issue is the specification of
N=2 finite-dimensional superalgebra in 2-dimensions
on which the topological twist will be performed.
The topological twist usually means the
mixing of the representation space of the internal symmetry
group of the supersymmetry with that of the symmetry group
with respect to the space-time,
i.e. spinor space. On manifolds,
the latter symmetry is local.
Consequently, the former symmetry must also be local.
The situation is allowed in the case of the
conformal supersymmetry alone.
On the contrary, the other types of the super-extended
algebra, that is, super-Poincar\'e or super-(anti-)de-Sitter,
must not be adapted for the present case
because it is not possible to deal with its
internal symmetry as a purely geometrical object
in contrast to the conformal case.
In the first place, the topologically twisted
super-Poincar\'e algebra is incomplete
from the geometrical view point
\cite{j.labastida&p.llatas1}\cite{a.galperin&o.ogievetsky1}.

What we are next interested in is the finite-dimensional
conformal superalgebras.
The finite-dimensional simple Lie superalgebras
are fully investigated\cite{superalgebra}
and all the finite-dimensional conformal superalgebras
in two dimensions are shown\cite{j.mccabe&b.velikson1}.
The 4-types of all the algebras must be eliminated from
the physical view point;$s_2$,
osp(2,1$\vert $N), su(1,1$\vert $1,1) and
d(1,2;$\alpha $)\cite{j.mccabe&b.velikson1}.
The 4-types are unsuitable also for the present case,
either because there exists
no anti-commutator of the supercharges
or because the supersymmetry is in the representations of integer spin.
We will soon understand the reason why in the forthcoming contexts.
After all, the possible finite-dimensional conformal
superalgebras in 2-dimentions
are then osp(N$\vert $2) (N$\geq $0), su(N$\vert $1,1) (N$\geq$2),
$f_4$ and $g_3$.
Moreover, it is the N=2 case that we are interested in.
In the case, the remaining card is osp(2$\vert $2) alone.
Therefore, osp(2$\vert $2) is the unique solution
for performing the topological twist on finite-dimensional
N=2 superalgebra in two dimensions.

The internal symmetry group of \Osp
in relation to $(2,2)$ supersymmetry is SO(2)$\otimes$SO(2), and
Osp(2$\vert$2) is required to be compact so that
its Cartan-Killing form is positive definite, while
the super Lie group Osp(2$\vert$2) is generally not compact.
\osp conformal superalgebra on which
the twisting operation will be made
to form a corresponding topological algebra
is then confined to the two dimensional
Lorentzian manifold with the local Lorentz metric
in the light-cone coordinates:
$g^{z\bar{z}}=g^{\bar{z}z}=-2,\enskip g_{z\bar{z}}=g_{\bar{z}z}
=-1/2$ and $g^{zz}=g^{\bar{z}\bar{z}}=
g_{zz}=g_{\bar{z}\bar{z}}=0$.
This (2,2) superalgebra contains two
types of complex Weyl spinorial charges
$Q,\enskip \bar{Q};\enskip S,\enskip \bar{S}$,
where ``$-$'' means the Dirac conjugation
$\bar{Q}=Q^{\dagger}\gamma^0$ in
which $\gamma^0=i\sigma^2$ and incidentally
$\gamma^1=\sigma^1,\enskip
\gamma^5=-\gamma^0\gamma^1=\sigma^3$,
or equivalently
$\gamma^z=\gamma^0+\gamma^1,\enskip
\gamma^{\bar{z}}=\gamma^0-\gamma^1$
in the light-cone coordinates.
These supercharges are two component spinors,
for example $Q=(Q_+,\enskip Q_-)^t$, where ``$+,-$'' mean spinor indices
describing ``left'' and  ``right'' moving
, respectively, with respect to the local Lorentz coordinates
$(z,\bar{z})$. These indices are raised and lowered
by a metric in spinor space
given by the charge conjugation matrix $C=\gamma^0$:
$\eta^{+-}=\eta_{-+}=-1,\quad \eta^{-+}=\eta_{+-}=1$ and
$\eta^{++}=\eta^{--}=\eta_{++}=\eta_{--}=0.$

We can leave out the conjugate
parts of the bra-ket bosonic relations
with respect to the complex supercharges of \osp as follows:
\[
[S,P_a ]= \gamma_a Q,\hspace{1.0cm}
[S,D]=-\dfrac{1}{2} S,\hspace{1.0cm}[S,M]=-\dfrac{1}{2} \gamma_5 S,
\]
\[
[Q,K_a ]=-\gamma_a S,\hspace{1.0cm}
[Q,D]=\dfrac{1}{2} Q,\hspace{1.0cm}[Q,M]=-\dfrac{1}{2} \gamma_5 Q,
\]
\[
[S,A]=-i\dfrac{1}{4} \gamma_5 S,\hspace{1.5cm}[S,V]=-i\dfrac{1}{4} S,
\] \be
[Q,A]=i\dfrac{1}{4} \gamma_5 Q,\hspace{1.5cm}[Q,V]=-i\dfrac{1}{4} Q.
\label{osp3}\ee
If we want to get these conjugate parts of eqs.(\ref{osp3}),
we must pay attention to
the fact that the representation of
the body so(2)$\oplus$sp(2)
of osp(2$\vert$2) are anti-Hermitian where the
anti-Hermitian character of the representation of sp(2) actually
leads to the positivity of the Cartan-Killing form of osp(2$\vert$2).

Ordinary (2,2) supersymmetry which is free from
the central charges is direct sum of (2,0) and (0,2)
and the corresponding part in \osp reads
\[
\{Q_+,\bar{Q}_+\}=iP_z,\hspace{0.9cm}\{S_+,\bar{S}_+\}=-iK_z,
\]
\be
\{Q_-,\bar{Q}_-\}=iP_{\bar{z}},
\hspace{0.9cm}\{S_-,\bar{S}_-\}=-iK_{\bar{z}}.
\label{osp1}\ee
While the super-extended conformal algebra
has no central charge,
there are mixing parts, instead,
in the relations
between the supercharges, and consequently
the decomposition mentioned above does not exist.
The mixing part of (2,0) and (0,2) in \osp is
\[
\{Q_+,\bar{S}_-\}=i(M-D)+2(A-V),
\]
\[
\{Q_-, \bar{S}_+ \}=i(M+D)+2(A+V),
\]
\[
\{\bar{Q}_+,S_-\}=i(M-D)-2(A-V),
\] \be
\{\bar{Q}_-,S_+ \}=i(M+D)-2(A+V).
\label{osp2}\ee
The property will play an important role in the forthcoming contexts.

The bosonic generators of \osp are as follows:
$P_a,\enskip K_a,\enskip M,\enskip D,\enskip A$ and $V$ are
translation, conformal-translation, Lorentz, Weyl,
chiral so(2) and internal so(2), respectively.
The finite two-dimensional conformal algebra
composed of these bosonic generators alone is
\[
[P_a,M]=\epsilon_{ab}P^b,
\hspace{1.5cm}[P_a,D]=P_a,\hspace{1.5cm}[K_a,D]=-K_a,
\]
\be
[K_a,M]=\epsilon_{ab}K^b,
\hspace{1.5cm}[K_a,P_b]=2(\epsilon_{ab}M-\delta_{ab}D),
\label{osp4}\ee
where $\epsilon$ are
$\epsilon^{z{\bar z}}=-\epsilon^{{\bar z}z}=-2,\enskip
\epsilon_{z{\bar z}}=-\epsilon_{{\bar z}z}=1/2$
and $\epsilon^{zz}=\epsilon^{\bar{z}\bar{z}}=
\epsilon_{zz}=\epsilon_{\bar{z}\bar{z}}=0$.

Lastly in the presentation of \osp algebra,
let us comment on the naming of the generators of \osp.
In sect.3 where a pure gauge theory of \osp on two
dimensional manifold will be considered,
the naming, for instance, $P$ as translation, is
perfectly formal.
If not, the general coordinate transformations must exist
in the system and then the theory may become empty
as well as the ordinary 2D conformal
supergravity theories\cite{supergravityconstraint}.\\

\befl \underline{ 2-2 \ Topological Twist} \efl
\hed We are now in a position to perform
topological twisting of the algebra. Topological twist
is usually a kind of mixing which
results in identification of the representation space of
internal symmetry group of N=2
supersymmetry with that of the local Lorentz group.
It is easy to perform
twisting of the algebra to get the topological algebra.
Most of all we have to do
is to replace $Q$, $\bar Q$, $S$, and $\bar S$ with
$Q^+$, $Q^-$, $S^+$, and $S^-$, respectively.
The indices ``$+,-$'' are raised and lowered with the same metric
as for the indices $\alpha ;\beta$ of $C_{\alpha \beta}$ and $Q_{\alpha}$.
That is, the complex Weyl spinors $\varphi_\alpha$,
$\bar{\varphi}_\alpha$ are substituted for
$\varphi_\alpha ^{\enskip +}$, $\varphi_\alpha^{\enskip -}$:
\be
\varphi_\alpha =\dfrac{i}{\sqrt{2}}\varphi_\alpha^{\enskip +},
\quad \bar{\varphi}_\alpha =
\dfrac{i}{\sqrt2} \varphi_\alpha ^{\enskip -}.
\label{topoto} \ee

The remaining manipulations are as follows. The fermionic charges
$Q^+=(Q_+^{\enskip +} ,Q_-^{\enskip +})^t$ have
become ( (0,0)-form , (0,1)-form ),
and $Q^-=(Q_+^{\enskip -} ,Q_-^{\enskip -})^t$
with ( (1,0)-form , (0,0)-form ),
{\it {idem}} $S^{\pm}$.  Then we have to
modify the definitions of local Lorentz $M$
and Weyl $D$ generators so that the four (0,0)-form fermionic generators
of supersymmetry have no charge with respect to
these two bosonic generators. We have put the representation
space accompanied with the internal symmetry
group SO(2)$\otimes$SO(2) upon
the space of spinor. The modified $M$, $D$
generators must be direct sums
with so(2)$\oplus$so(2)
generators $V$ and  $A$, respectively. The solution to this
constraint resolves into
\be
\begin{array}{cc}
\tilde{M} =M+2iV, & \tilde{D} =D+2iA.
\end{array}
\label{c}
\ee
These modified generators then satisfy the following relations:
\be
\begin{array}{cc}
[ \triangle_{\pm}^{\enskip \pm}, \tilde{M} ]=0,
& [ \triangle_{\pm}^{\enskip \pm}, \tilde{D} ]=0,
\end{array}
\label{d}
\ee
where $\triangle$ means both $Q$ and $S$.

There appear some problems about the closure
of the modified algebra, however.
The generators $A$ and $ V $ have been
put upon $D$ and $M$, respectively
and the modified algebra which contains $\tilde{M}$ and $\tilde{D}$
must not contain $A$ and $ V $.
In fact, the modified algebra contains subtle relations:
\[
\{Q_-^{\enskip +}, S_+^{\enskip -}\}
=i(\tilde{M}+\tilde{D})-4i(A+V),
\]
\be
\{Q_+^{\enskip -}, S_-^{\enskip +}\}
=i(\tilde{M}-\tilde{D})+4i(A-V).
\label{e}\ee

We can avoid the above relations (\ref{e}) as in the followings.
Here it is necessary to omit
another generators with regard to eqs.(\ref{e}),
if this modified algebra still obeys the closure
property for the generators of the gauge symmetry.
In this point of view, the four fermionic generators
$Q_+^{\enskip -}$, $Q_-^{\enskip +}$,
$S_+^{\enskip -}$ and $S_-^{\enskip +}$
do not induce the gauge transformations generated by both
$i(\tilde{M}+\tilde{D})-4i(A+V)$ and $i(\tilde{M}-\tilde{D})+4i(A-V)$.
There are two alternatives, that is,
the case in which the left chiral charges
$Q_+^{\enskip -}, S_+^{\enskip -}, P_z,$ and $K_z$
vanish, or the case in which the right chiral charges
$Q_-^{\enskip +}, S_-^{\enskip +}, P_{\bar{z}},$ and
$K_{\bar{z}}$ vanish,
without any compensation procedure,
that is, all gauge fields and parameters of these four generators are
assured to vanish.
The second case is adapted here.
In sect.4, it will be shown that a moduli space derived from
either case is reduced to that associated with
an intersection part of both cases.

The twisting procedure is explained in terms of the gauge fields of the
corresponding symmetry \osp. Let us introduce the gauge
field ${\bf a}$ which is Lie superalgebra-valued
1-form of \osp in the form:
\begin{eqnarray}
{\bf a}_\mu = e_\mu ^a P_a & +&  f_\mu ^a K_a  +
\omega_\mu M + b_\mu D \nonumber \\
  & + & a_\mu A + v_\mu V +\bar{\psi}_\mu Q + \bar{Q} \psi_\mu +
\bar{\phi}_{\mu} S + \bar{S} \phi_\mu,
\end{eqnarray}
as well as transformation parameter ${\bf {\tau}}$ defined by
\begin{eqnarray}
{\bf {\tau}} = \xi_P ^a P_a & + & \xi_K ^a K_a  +
\lambda_l M + \lambda_d D \nonumber \\
& + & \theta_a A + \theta_v V +
\bar{\varepsilon} Q + \bar{Q} \varepsilon
+ \bar{\kappa} S + \bar{S} \kappa.
\label{para}\end{eqnarray}
Using the gauge fields and parameters, the above mentioned
topological twist and additional manipulations can be described as follows:
eqs.(\ref{c}) mean
\be
v_\mu=2i\omega_\mu, \quad a_\mu = 2ib_\mu,
\label{modcon} \ee
and elimination of the generators $Q_-^{\enskip +}$,
$S_-^{\enskip +}$, $P_{\bar{z}}$
and $K_{\bar{z}}$ means
\begin{eqnarray} \vth
\phi_{\mu +}^{\quad -} = 0 = \psi_{\mu +}^{\quad -},&\quad &
\kappa_+^{\quad -} = 0 = \varepsilon_+^{\quad -},\nonumber \\ \vth
e_\mu ^{\bar{z}} = 0 = f_\mu ^{\bar{z}},&\quad &
\xi_P ^{\bar{z}} = 0 = \xi_K ^{\bar{z}}.
\label{figcon} \end{eqnarray}

Under the conditions we are led to
\be
\delta \phi_{\mu +}^{\quad +} \sim \delta \phi_{\mu -}^{\quad -}, \quad
\delta \psi_{\mu +}^{\quad +} \sim \delta \psi_{\mu -}^{\quad -}.
\label{equiv} \ee
Accordingly, we have the following identifications:
\[ \psi_{\mu +}^{\quad +} = -\psi_{\mu -}^{\quad -} \equiv -\psi_\mu,
\quad \phi_{\mu +}^{\quad +} = -
\phi_{\mu -}^{\quad -} \equiv -\phi_\mu,\]
\be
\varepsilon_+^{\enskip +} = -\varepsilon_-^{\enskip -}
\equiv -\varepsilon,\quad
\kappa_+^{\enskip +} = -\kappa_-^{\enskip -} \equiv -\kappa,
\label{id} \ee
which read without loss of generality
\be
Q \equiv Q_+^{\enskip +} + Q_-^{\enskip -},\quad S
\equiv S_+^{\enskip +} + S_-^{\enskip -}.
\label{cha} \ee
Taking into account all these additional conditions
with respect to the topological twist on
the original \osp,
we get the gauge connection ${\bf a}$:
\begin{eqnarray}
{\bf a}_{\mu}=e_{\mu}^z P_z + f_{\mu}^z K_z &+&
\omega_{\mu} \tilde{M} + b_{\mu} \tilde{D} \nonumber\\
 &-& \dfrac{1}{2} (Q_+^{\enskip -}\psi_{\mu -}^{\quad +} +
S_+^{\enskip -}\phi_{\mu -}^{\quad +}
+ \psi_{\mu} Q + \phi_{\mu} S),\label{fineqc}\end{eqnarray}
and transformation parameter ${\bf {\tau}}$:
\begin{eqnarray}
{\bf {\tau}}= \xi_P^z P_z + \xi_K^z K_z &+&
\lambda_l \tilde{M} + \lambda_d \tilde{D}\nonumber \\
&-& \dfrac{1}{2} (Q_+^{\enskip -}\varepsilon_-^{\enskip +}
+ S_+^{\enskip -}\kappa_-^{\enskip +}
+ \varepsilon Q +\kappa S),\label{fineq} \end{eqnarray}
respectively.

After all, the generators in
eqs.(\ref{fineqc}) (\ref{fineq}) obey the following relations:
\[ \begin{array}{lll}
[S,P_z]= Q_+^{\enskip -}, &
[Q_+^{\enskip -},\tilde{D}]=Q_+^{\enskip -},&
[Q_+^{\enskip -},\tilde{M}]=-Q_+^{\enskip -},
\end{array} \]
\[ \begin{array}{lll}
[Q,K_z]=-S_+^{\enskip -}, &
[S_+^{\enskip -},\tilde{D}]=-S_+^{\enskip -},&
[S_+^{\enskip -},\tilde{M}]=-S_+^{\enskip -},
\end{array} \]
\[ \begin{array}{lll}
\{ Q,Q_+^{\enskip -} \} =-2iP_z,&
\{S,S_+^{\enskip -} \} =2iK_z, &
\{ Q,S \} = -4i\tilde{M},
\end{array} \]
\[ \begin{array}{lll}
[P_z, \tilde{M}]=-P_z, & [P_z, \tilde{D}]=P_z, & \quad
\end{array} \]
\be \begin{array}{lll}
 [K_z, \tilde{M}]=-K_z, & [K_z, \tilde{D}]=-K_z, & \quad
\end{array}
\label{topoall}\ee
and the gauge connections (\ref{fineqc}) satisfy
the following transformation rules:
\[
\delta\psi_\mu=\partial_\mu \varepsilon,
\] \[
\delta \phi_\mu =\partial_\mu \kappa ,
\] \[
\delta \psi_{\mu -}^{\quad +} = {\cal D}_{\mu}\varepsilon_-^{\enskip +}
  + \xi_P^z \phi_\mu + ( \lambda_l - \lambda_d )
  \psi_{\mu -}^{\quad +} - e_\mu^z \kappa ,
\] \[
\delta \phi_{\mu -}^{\quad +} = {\cal D}_{\mu}\kappa_-^{\enskip +} - \xi_K^z
  \psi_\mu + ( \lambda_l + \lambda_d)
  \phi_{\mu -}^{\quad +} + f_\mu^z \varepsilon ,
\] \[
\delta e_\mu^z ={\cal D}_\mu \xi_P^z +
  (\lambda_l - \lambda_d) e_\mu^z
  -\dfrac{i}{4} (\varepsilon \psi_{\mu -}^{\quad +}-
  \psi_\mu \varepsilon_-^{\enskip +}) ,
\] \[
\delta f_\mu^z={\cal D}_\mu \xi_K^z +
  (\lambda_l + \lambda_d) f_\mu^z +\dfrac{i}{4}
  (\kappa \phi_{\mu -}^{\quad +} - \phi_\mu \kappa_-^{\enskip +}) ,
\] \[
\delta \omega_\mu = \partial_{\mu}\lambda_l
  +\dfrac{i}{4} (\kappa \psi_{\mu} - \varepsilon \phi_{\mu}),
\] \be
\delta b_\mu = \partial_\mu \lambda_d,
\label{gaugeal} \ee
where
\[ \begin{array}{ll}
{\cal D}_\mu \varepsilon_-^{\enskip +}=
  (\partial_\mu -\omega_\mu +b_\mu)\varepsilon_-^{\enskip +}, &
{\cal D}_\mu \kappa_-^{\enskip +}=
  (\partial_\mu - \omega_\mu - b_\mu)\kappa_-^{\enskip +},
\end{array} \]
\be \begin{array}{ll}
{\cal D}_\mu \xi_P^z =
  (\partial_\mu - \omega_\mu + b_\mu)\xi_P^z, &
{\cal D}_\mu \xi_K^z =
  (\partial_\mu - \omega_\mu - b_\mu)\xi_K^z,
\end{array}
\ee
The field strengths in relation to the discarded
right chiral charges $Q_-^{\enskip +}$,
$S_-^{\enskip +}$, $P_{\bar{z}}$ and $K_{\bar{z}}$
all vanish as expected.
The resultant algebra (\ref{topoall}) can be
referred to as the topological algebra\cite{j.labastida&p.llatas1}.

SO(2)$\otimes$SO(2) symmetry still remains as global internal
symmetry whose charge is the so-called ghost-number,
the generators of which are defined by $G\equiv 2i(A-V)$,
$\tilde{G}\equiv 2i(A+V)$.
Here $G$ and $\tilde{G}$ satisfy the following relations:
\[ \begin{array}{ll}
[G,Q_+^{\enskip +}]=Q_+^{\enskip +}, &
[G,S_+^{\enskip +}]=-S_+^{\enskip +},
\end{array} \]
\[ \begin{array}{ll}
[G,Q_+^{\enskip -}]=-Q_+^{\enskip -}, &
[\tilde{G},S_+^{\enskip -}]=S_+^{\enskip -},
\end{array} \]
\be \begin{array}{cc}
[\tilde{G},Q_-^{\enskip -}]=Q_-^{\enskip -}, &
[\tilde{G},S_-^{\enskip -}]=-S_-^{\enskip -},
\end{array}
\label{i}
\ee
where the other combinations are trivial.
As a consequence of eqs.(\ref{i}), indeed,
it is natural to regard these generators $G$, $\tilde{G}$
as the ghost number operators.
$Q_{\pm}^{\enskip \pm}$ and $S_+^{\enskip -}$ increase the ghost
number by one unit, while $Q_+^{\enskip -}$
and $S_{\pm}^{\enskip \pm}$ decrease it by the
same quantity.
The assignment is consistent with the
relations (\ref{topoall}) (\ref{gaugeal}).

In preparation for the forthcoming contexts,
next, the description
of the coordinate indices in the relations (\ref{topoall})
must be simplified.
First of all, the local Lorentz coordinates are substituted for
spinor indices of the supercharges as follows:
\be\begin{array}{ccc}
Q_+^{\enskip -}=2Q_z, &\quad
S_+^{\enskip -}=2S_z,
\end{array}\label{index}\ee
from the following relations of fermionic field $\varphi$:
\be
\begin{array}{ccccccc}
 \varphi_\alpha^{\enskip \beta}  &
 =  &
 \varphi_a (\gamma^a)_\alpha^{\enskip \beta}  &
 =  &
 \begin{array}{c}
           \scriptstyle + \\ -
        \end{array}
          \begin{array}[b]{c}
              \begin{array}{cc}
                \scriptstyle + & -
              \end{array}\\
              \left(
                    \begin{array}{cc}
                    0 & 2 \\
                    0 & 0
                    \end{array}
              \right)
         \end{array} \varphi_z  &
 +  &
 \begin{array}{c}
            \scriptstyle + \\ -
        \end{array}
          \begin{array}[b]{c}
             \begin{array}{cc}
               \scriptstyle + & -
             \end{array}\\
             \left(
                  \begin{array}{cc}
                    0 & 0 \\
                   -2 & 0
                  \end{array}
             \right)
         \end{array} \varphi_{\bar z} ,
\end{array}\ee \\
where $\alpha$($\beta$) is ``$+, -$''
and ``a'' means ``$z$, $\bar{z}$''.
The above supercharges are further substitutable as follows:
\be
Q_z = q_z Q_c, \quad S_z = s_z S_c,
\ee
where $q_z$ and $s_z$, carring the chiral idex; $z$,
commute all generators
in the algebra (\ref{topoall})
and ``c'' means the ``chiral''.
$q_z$ and $s_z$ are left chiral components of real
vectors $q=(q_z, q_{\bar z})^t$, $s=(s_z, s_{\bar z})^t$, respectively.
Therefore, $Q_c$ and $S_c$ are real generators.
We then obtain a different description of the
topological algebra (\ref{topoall}):
\[ \begin{array}{lll}
[S, q^z P_z]=2q Q_c, &
[Q_c, \tilde{D}]=Q_c, &
[Q_c, \tilde{M}]=-Q_c,
\end{array} \]
\[ \begin{array}{lll}
[Q, s^z K_z]=-2sS_c, &
[S_c, \tilde{D}]=-S_c, &
[S_c, \tilde{M}]=-S_c,
\end{array} \]
\[ \begin{array}{lll}
\{ Q, Q_c \} = - \dfrac{i}{q} q^z P_z, &
\{ S, S_c \} = \dfrac{i}{s} s^z K_z, &
\{ Q, S \} =-4i \tilde{M},
\end{array} \]
\[ \begin{array}{ll}
[P_z, \tilde{M}]=-P_z, &
[P_z, \tilde{D}]=P_z,
\end{array} \]
\be \begin{array}{ll}
[K_z, \tilde{M}]=-K_z, &
[K_z, \tilde{D}]=-K_z,
\end{array}
\label{topoal}\ee
where $q=q^z q_z$ and $s=s^z s_z$.

We must note that the four generators
$q^zP_z$, $s^zK_z$, $Q_c$ and $S_c$
still behave as holomorphic one forms
because the real vectors $q$, $s$ commute all generators
in the algebra (\ref{topoal}) and then
the commutation relation with $\tilde{M}$
is still retained.
It is a matter of course that,
if 2-manifold $\mbox{M}^2$ is Hermitian with no boundary,
four genarators $q^zP_z$, $s^zK_z$, $Q_c$ and $S_c$
could behave as zero forms, that is, they
commute $\tilde{M}$,
regarding $q^z(s^z)$ as the ordinary adjoint Dolbeault operator;
$\partial^\dagger(=-\ast \bar{\partial}\ast)$
which satisfies the relation
$[\tilde{M}, \, \partial^\dagger ]=-\partial^\dagger$.
We must note that the scale dimensions
can not be wiped out
through $\partial^\dagger$, however.\\

\section{\rm Vanishing Noether Current}
\befl \underline{ 3-1 \ Brief Sketch of TFT} \efl
\hed In the previous section,
the twisted \osp algebra (\ref{topoal}) has been obtained.
We will show that the algebra (\ref{topoal}) has the
TFT's Lagrangians and the moduli
space associated with the algebra can be derived.
Our principal concern is now reduced to building up
the TFT Lagrangian and quantizing it.
Let us give a brief sketch of
TFT\cite{d.birmingham&m.blau&m.rakowski&g.thompson1}
in preperation for the following discussions.
The theory has some fermionic operator
${\cal Q}$ of nilpotency and the Lagrangian ${\cal L}$ is
${\cal Q}$-exact: ${\cal L}=\{ {\cal Q}, \star \}$.
The ${\cal Q}$-exact form of $\cal L$
shows that the energy-stress tensor of the
Lagrangian is also ${\cal Q}$-exact.
This is non-trivial because of no informaton whether
$\delta /\delta g_{\mu \nu} $ and ${\cal Q}$ are commutative or not.
All known examples are in the case, or rather,
due to the ${\cal Q}$-exact form of the stress-energy tensor,
it is possible to discuss the topological invariat correlation functions.

The TFT Lagrangian is allowed
to have the gauge symmetries, in which the cohomological nature of TFT
turns out to the equivariant cohomology:
$ {\cal Q}^2=\tau_{\phi}$, where $\tau_{\phi}$
means the gauge transformation with the parameter $\phi$.
The freedom of the gauge symmetry can be fixed by using
BRST method as usual before the quantization of this system.
In TFT this freedom may be automatically fixed
after the quantization through the path-integral,
i.e. the reduction to some moduli space.

The correlation functions
are independent of a coupling factor,
because ${\cal L}$ is ${\cal Q}$-exact
The zero limit of the coupling factor then
induces the leading contribution of the path-integral with ${\cal L} =0$.
This configuration of the fields is associated with a moduli space.
As a consequence, TFT makes a local field theory
in the sense of a finite integration
on the moduli space after quantization.

It is necessary to mention that
the behavior of the path-integral measure is crucial
in the case of full construction
of TFT\cite{e.witten2}, e.g. in studies of the mirror
symmetry\cite{e.witten4}\cite{p.aspinwall&d.morrison1}.
The fact that, due to the coupling factor independence of
correlation functions, the configuration of
the theory could be reduced to the classical
one, i.e. zero mode configuration, does not
necessarily lead us to the conclusion
that the theory could be regarded perfectly
as classical in itself.
The non-zero mode contribution must
not be disregarded in the
case of full construction of TFT.
Anyhow, if we want to know the classical configuration
alone, it is not necessary to take into account
the non-zero mode as a quantum effect.

Due to the zero-mode contribution
associated with the detailes of the moduli space,
there is a close relation between the behavior of
the measure and the triviality of the path-integral.
The triviality depends on
the characterestic of the TFT obsrevables
which satisfy at least the following conditions:
the gauge invariance, the metric idependence
and the elements of ${\cal Q}$-cohomology group
(rigorously speaking, the gauge invariance and
the metric independence, modulo ${\cal Q}$-exact).
In the path-integration under the limit,
the integral measure must be reduced to
that on the classical configuration
and the fermionic number anomaly
which corresponds to the local dimension of
the classical configuration comes manifestly to the measure.
Therefore, the non-trivial correlation functions must be given
by the observables with the same fermionic number
as the measure's.
Consequently, we must need some proper obsevables,
in other words, an assurance that the correlation
functions are non-trivial.
At this stage, it may not be possible
to present such an assurance.
In the next section, it will be shown that
the path-integral in the present case is non-trivial
through discussion on a moduli space.\\

\befl \underline{ 3-2 \ Algebraic Lagrangian} \efl
\hed Let us turn attention to building up the TFT's Lagrangian.
We can find the specific relations in the algebra (\ref{topoal}):
\be
\begin{array}{ccc}
\{ Q, Q_c \} = \mbox{$ - \dfrac{i}{q}$} q^z P_z,  &
\{ S, S_c \} = \mbox{$\dfrac{i}{s}$} s^z K_z,  &
\{Q, S \}= -4i \tilde{M} .
\end{array}
\label{lagrangians3}\ee
The above three relations are in the form of ${\cal Q}$-exact
with the ghost number zero,
and the above fermionic operators are nilpotent.
Therefore, it may be possible to regard
the relations (\ref{lagrangians3}) as
{\it algebraic} Lagrangians of TFT:
\be
\begin{array}{ccc}
{\cal L}_Q = \{ Q, Q_c \}, &
{\cal L}_S = \{ S, S_c \}, &
{\cal L}_{QS} = \{ Q, S \}.
\end{array}
\label{def-of-lagrangians}\ee
The identifications (\ref{def-of-lagrangians})
do not seem natural from the relativistic
field theoretical view point
because ${\cal L}_Q$ and ${\cal L}_S$ have
non-zero spins and scaling dimensions.
Intuitively speaking, ${\cal L}_Q$ and ${\cal L}_S$ are not neutral.
We will comment on plausibility and uniqueness
of the identificaltions
(\ref{def-of-lagrangians}) in the
last two paragraphs of this subsection 3-2.

The three Lagrangians
(\ref{def-of-lagrangians}) are real
and must be defined in two dimensional
manifolds with boundary.
For instance, ${\cal L}_{QS}$ is described
as two dimensional integration:
\be
{\cal L}_{QS} = \int_{\partial \mbox{\scriptsize M}^2}J_{ \{ Q, S \} }^0 =
\int_{\mbox{\scriptsize M}^2}dJ_{ \{ Q, S \} }^0.
\ee
The existence of boundary is in the way for the following discussions.
Notwithstanding, we can construct the quantum theories associated
with the Lagrangians (\ref{def-of-lagrangians})
on manifolds without boundary.
It is well known that the path-integral on manifold
with boundary describes some states on boundary,
which are also topological invariants\cite{e.witten1},
and consequently we can make the formulation
on manifold without boundary by means of the inner
product of the path-integrals as ``in'' and
``out'' states. Hereafter we then suppose that the manifold
of the theory will without boundary
\cite{m.blau&g.thompson1}\cite{e.witten5}.

Each system of the three Lagrangians (\ref{def-of-lagrangians})
has the manifest BRST-like fermionic symmetries as follows:
The system of ${\cal L}_Q$ has $Q$- and $Q_c$-symmetries,
the system of ${\cal L}_S$ has $S$- and $S_c$-symmetries
and the system of ${\cal L}_{QS}$ has $Q$- and $S$-symmetries.
The energy-stress tensor of each system is
a liner-combination of exact-forms of the two fermionic generators
which the system has.
For instence, the energy-stress tensor of ${\cal L}_{QS}$ is
\be
T_{\mu \nu} = \{ Q, \delta S /\delta g_{\mu \nu} \} +
                        \{ S, \delta Q /\delta g_{\mu \nu} \}.
\label{est}\ee
The equation (\ref{est}) is based on the fact that
we have no information whether $Q$ and $S$ are dependent on
the metric $g_{\mu \nu}$ or not.
It is well known that the correlation functions of ${\cal Q}$-exact
operators are trivial in TFT\cite{e.witten1}:
\be
\langle \enskip \mbox{${\cal Q}$-exact } \enskip \rangle = 0,
\label{q-exact}\ee
where $\langle \cdot \cdot \cdot \rangle$ means
the path-integration.
Therefore, the $Q$- and $S$-exact form of the energy-stress tensor
guarantees that the correlation functions
of any non-trivial TFT's observables ${\cal O}$ are independent on
the metric $g_{\mu \nu}$:
\begin{eqnarray}
\dfrac{\delta}{\delta g_{\mu \nu}} \langle \enskip {\cal O} \enskip
\rangle & = & \langle \enskip {\cal O} T_{\mu \nu} \enskip \rangle
\nonumber\\
& = & \langle \enskip {\cal O} \{ Q \mbox{-exact} \} \enskip \rangle +
          \langle \enskip {\cal O} \{ S \mbox{-exact} \} \enskip \rangle
\nonumber\\
& = & \langle \enskip Q \mbox{-exact} \enskip \rangle +
          \langle \enskip S \mbox{-exact} \enskip \rangle
\nonumber\\
& =& 0.
\end{eqnarray}
Consequently, we regard $Q$, $S$, $Q_c$ and $S_c$
as the topological symmetry genetators
in each of the three systems
(\ref{def-of-lagrangians}).

At the classical level, the transformation rules of the three
Lagrangians (\ref{def-of-lagrangians})
under the symmetry
generated by the topological
algebra (\ref{topoal}) are as follows:
\[ \begin{array}{ccc}
\delta_{A^\star} {\cal L}_Q = 0 &
\mbox{up to} &
\mbox{$ Q,\enskip Q_c $ -exact terms,}
\end{array} \]
\[ \begin{array}{ccc}
\delta_{A^\star} {\cal L}_S = 0  &
\mbox{up to} &
\mbox{$ S,\enskip S_c $ -exact terms,}
\end{array} \]
\be \begin{array}{ccc}
\delta_{A^\star} {\cal L}_{QS} =0  &
\mbox{up to} &
\mbox{$ Q,\enskip S $ -exact terms,}
\end{array}
\label{law}\nonumber\ee
where $A^\star$ denotes a set of the generators
of the algebra (\ref{topoal}):
\be
A^\star = P_z,K_z,\tilde{M},\tilde{D},Q,Q_c,S,S_c .
\label{info}
\ee
The relations (\ref{law}) show that, at the quantum level,
every system is invariant under the symmetry which
is generated by $A^\star$ because of (\ref{q-exact}).

It is necessary to comment on the above identifications
(\ref{def-of-lagrangians}).
In the relativistic field theories, the Lagrangians must
be invariant under the local Lorentz transformations.
Unfortunately, ${\cal L}_Q$ and ${\cal L}_S$ in
the definitions (\ref{def-of-lagrangians})
are not invariant under the Lorentz transformations,
so that  ${\cal L}_Q$ and ${\cal L}_S$ have integer spin 1.
Moreover, they have non-zero scaling dimensions.
These facts will lead us to conclude that
${\cal L}_Q$ and ${\cal L}_S$ are not suitable
for Lagrangians from the conventional field theoretical view point.
As mentioned above,
the path-integrals with such Lagrangians are surely invariant
under the local Lorentz and Weyl transformations.
Therefore, let us regard ${\cal L}_Q$ and ${\cal L}_S$
as Lagrangians of TFT.

There is a question whether the other candidates for Lagrangian
exist or not. In the algebra (\ref{topoal}),
the remaining relations to be regarded as Lagrangian
are all in the same form:
\be
[{\cal F}_{\mbox{\scriptsize fermi}},\;
{\cal B}_{\mbox{\scriptsize bose}}]
=\hat{{\cal F}}_{\mbox{\scriptsize fermi}}.
\label{special}\ee
While the above commutation relation
(\ref{special}) is indeed ${\cal Q}$-exact, it is fermiomic.
{}From the physical view point,
it is not possible to regard such a fermionic relation
as Lagrangian.
Clearly, if so, the perturbation could be up to the first order.
Moreover, the generators of Poincar\'e group,
which are obtained through Lagrangian with
Poincar\'e group invariace,
would be composed of anti-commutators alone and then
the crucial relation between the symmetry and the conservation law
would be nothing.
The Lagrangian formalism (or more generally the mechanics)
would break down.
Therefore, such a quasi-Lagrangian which is fermionic
is by no means accepted as the physical Lagransian.
Notwithstanding, there is no reason to disregard
the fermionic quasi-Lagrangian
if we are allowed formally to regard the quasi-Lagrangian
as mere exponent of the Boltzman weight factor
in the partition function.
What we are interested in is whether,
under the zero coupling limit,
the classical configuration could be induced or not.

Let us now evaluate the possibility of the
fermionic quasi-Lagrangian.
In the case of ${\cal B}_{\mbox{\scriptsize bose}}=\tilde{M}$ or $\tilde{D}$,
we see ${\cal F}_{\mbox{\scriptsize fermi}}=
\hat{{\cal F}}_{\mbox{\scriptsize fermi}}$, that is, the transformation
\be
{\cal B}_{\mbox{\scriptsize bose}}:
{\cal F}_{\mbox{\scriptsize fermi}} \enskip
\mapsto \enskip {\cal F}_{\mbox{\scriptsize fermi}}
\label{map}\ee
is identity endomorphism and then
the weight factor of the path-integral becomes
the polynomial of ${\cal L}$-exact factors:
\begin{eqnarray}
\mbox{exp}( \dfrac{1}{\alpha} {\cal L})
& = & 1+\dfrac{1}{\alpha}{\cal L} \nonumber
\\
& = & 1+\dfrac{1}{\alpha}[{\cal L},\;
      {\cal B}_{\mbox{\scriptsize bose}}]
,
\end{eqnarray}
where $\alpha$ denotes a coupling factor.
Therefore, the path-integral with Lagrangian
which is the form $[{\cal F}_{\mbox{\scriptsize fermi}},\;
{\cal B}_{\mbox{\scriptsize bose}}]=
{\cal F}_{\mbox{\scriptsize fermi}}$
 yields infinite volume alone and the fact indicates that
the path-integral is trivial.
The limitation of the coupling factor turns out to be meaningless.

The remaining case is ${\cal B}_{\mbox{\scriptsize bose}}=P$ or $K$.
If we regard the relation (\ref{special}) with
${\cal B}_{\mbox{\scriptsize bose}}=P$ or $K$ as Lagrangians ${\cal L}$,
then ${\cal L}$ is invariant under the transformation
generated by ${\cal L}$ itself:
\be
\delta_{\cal L}{\cal L}=0.
\label{ll}\ee
In this case, ${\cal L}$ is transformed into itself by $\tilde{M}$
and $\tilde{D}$, that is, ${\cal L}$ is ${\cal L}$-exact:
\be
{\cal L}=[ {\cal L},\; {\cal B}_{\mbox{\scriptsize bose}} ],
\label{endmor}\ee
where ${\cal B}_{\mbox{\scriptsize bose}}=\tilde{M}$ or $\tilde{D}$.
The above two facts, that is,
the Lagrangian ${\cal L}$ is invariant under the
transformation generated by ${\cal L}$ itself (\ref{ll})
and ${\cal L}$ is ${\cal L}$-exact (\ref{endmor}),
indicate that the path-integral with the Lagrangian ${\cal L}$
is also trivial because of the same reason as in the above case of
${\cal B}_{\mbox{\scriptsize bose}}=
\tilde{M}$ or $\tilde{D}$.
Consequently, the Lagrangian candidates which yield non-trivial
path-integrals are ${\cal L}_{Q}$, ${\cal L}_{S}$
and ${\cal L}_{QS}$ alone.\\

\befl \underline{ 3-3 \ Reduction to Moduli Space} \efl
\hed We will show that a moduli space associated with the
algebra (\ref{topoal}) can be derived
just by focusing on the Lagrangian ${\cal L}_{QS}$
through a gauge system of \osp \\
algebra. The same result could be obtained in the
case of using the other Lagrangians ${\cal L}_Q$ or ${\cal L}_S$.

As mentioned above, the configuration of the system results in a
corresponding moduli space after the quantization in TFT.
The reduction to a moduli space is a result of the weak coupling limit.
Under the limit, the leading contribution could be
given by zero mode, that is, the classical configuration
which makes the Lagrangian vanish.
We now suppose that
there exists some proper observable
which guarantee the non-triviality of the path-integral.
Let us start with a vanishing Lagrangian condition.
Therefore, we obtain
\be
\ { \cal L } =  \{ Q,S \} = -4i{\tilde{M}} = 0.
\label{k}
\ee
There exists a system $\Gamma$ of the gauge fields
(\ref{fineqc}) of the topological algebra (\ref{topoal}).
Let a Noether current which generates
${\tilde M}$ be $J_0^{\tilde M}$.
{}From the condition (\ref{k}),
$dJ_0^{\tilde M}=0$ on 2-manifold
$\mbox{M}^2$ without boundary can be derived.
The condition means $J_0^{\tilde M}$ is constant
on $\mbox{M}^2$ for arbitrary 2D metric.
Moreover, it is necessary to estimate the behavior of
the path-integral defined on 2-manifold
with boundary under the zero coupling limit.
As mentioned above, a path-integral on manifold
with boundary can be regarded as a functional on boundary:
\be
Z_{\mbox{\scriptsize D}_i}[ \varphi ]=\int_{\psi
\vert_{\partial \mbox{\tiny M}^2}
=\varphi} {\cal D}\psi \: e^{-S( \psi ) },\enskip
(i=1,2),
\label{functor1}\ee
where $\mbox{M}^2=
\mbox{D}_1 \cup_{\partial
\mbox{\scriptsize M}^2}\mbox{D}_2$.
The boundary condition $\psi
\vert_{\partial \mbox{\scriptsize M}^2}=\varphi$
could be confined in a delta function with a proper periodicity.
The coupling factor independece of
$Z_{\mbox{\scriptsize D}_i}[ \varphi ]$
is evident because the invariance
of $Z_{\mbox{\scriptsize D}_i}[ \varphi ]$
under the gauge transformations by $Q$ and $S$
holds\cite{m.blau&g.thompson1}.
Therefore, we see that
\begin{eqnarray}
Z_{\mbox{\scriptsize D}_i}[ \varphi ] ({\cal O}) & = &
\int  {\cal D} \psi\: e^{ -S(\psi ) }\:
\delta_p ( \psi \vert_{\partial
\mbox{\scriptsize M}^2} -\varphi)\: {\cal O} \nonumber\\
& = & 0,
\label{functor2}
\end{eqnarray}
where ${\cal O} = Q$- or $S$-exact and $\delta_p$ denotes a
delta function with some proper periodicity.
The characterestic (\ref{functor2}) guarantees the coupling
factor independence of
$Z_{\mbox{\scriptsize D}_i}[ \varphi ] $
in the similar way
to the case of no-boundary.
Therefore, the zero coupling limit would induce
the classical configuration which makes
the Lagrangian vanish and then the condition
$J_0^{\tilde M} =0$ on boudary holds
because the Lagrangian could be described as
1-dimensional integration on
$\partial\mbox{M}^2$;
$ {\cal L}=\int_{\partial\mbox{\scriptsize M}^2}J_0^{\tilde M}$.
Consequently, the condition (\ref{k}) is reduced to
$J_0^{\tilde{M}}=0$ on 2-manifold without boundary.
We then define a sub-configuration
$\Gamma_s$ which satisfies $J_0^{\tilde{M}}=0$.
The constraint $J_0^{\tilde{M}} =0$ yields
the following reduction of the configuration:
\be
\Gamma \quad \Longrightarrow \quad \Gamma_s.
\label{l}
\ee

Let us confine ourselves to investigation of the physical meaning
of the reduction mentioned above just through the
Noether current which is composed of the connections of
original Osp(2$\vert$2)$\otimes$Osp(2$\vert$2) ($\equiv{\cal G}$).
To this aim, we will consider
a pure gauge theory of \osp, not supergravity theory.
Therefore, the naming of the generators of \osp,
which is shown in Sect.2,
is perfectly formal.
If not, that is, the naming is meaningful,
the general coordinate transformations must
be induced, and then the theory becomes empty as in the case of
the ordinary 2-dimensional conformal
supergravity gauge theory.

Let us start with considering the Yang-Mills action on
2-dimensional manifold without boundary:
\be
{\cal L}_{\mbox{\scriptsize YM}_2}=
\int_{\mbox{\scriptsize M}^2} |R_{\cal G}|^2 \ast 1,
\label{ym} \ee
where $\ast$ is Hodge star operator and
$R_{\cal G}$ is a field strength 2-form :$R_{\cal G}=R^A B I_{AB}$
in which $I_{AB}$ is
the Cartan-Killing matrix on the Lie algebra of ${\cal G}$.
Here the norm $|R_{\cal G}|^2$ has been
obtained by using the metric on $M^2$
and $I_{AB}$.
It is of interest to argue that the integrand
of eq.(\ref{ym}) can be rewritten as
$R^A \wedge \ast R^B I_{AB}$, together
with the volume form of the metric  $\ast 1$.
It is a matter of course that eq.(\ref{ym}) is invariant with respect to
the gauge symmetry ${\cal G}$ and has no general coordinate invariance.
The time component of the Noether current
in association with the symmetry generated by
$M$ then turns out to be
\be
J_M^0 = \dfrac{\partial {\cal L}_{\mbox{\scriptsize YM}_2}}
{\partial(\partial_0{\bf a}_{\mu}^A)}
G_M^A({\bf a})_{\mu}=R^BI_{AB}G_M^A({\bf a})_{\mu},
\label{current} \ee
where $G_M^A({\bf a})_{\mu}$ is defined by
\be
\delta_M{\bf a}_{\mu}^A=\lambda_M
G_M^A({\bf a})_{\mu},
\label{gg}\ee
${\bf a}$ is general form of the gauge connections
and $R^A=R_{01}^A$.
We can now add the optional field strength
components to the original current (\ref{current})
owing to the ambiguity of the Noether current.
We are free to choose the additional term:
\be
\partial_{\alpha}(\theta^AR^{A \alpha\beta}),
\label{add}\ee
where $\theta^A(x)$ 0-form is arbitrary function which supplements
the characteristics of $J_M^0$
with respect to the paired field strength $R^A$.
Here note that the summation convention for repeated
indices does not apply to indices of $\theta$ and
of exponent of statistical factor $(-)$
in the following equations.
It is a matter of course that the conventional
sum rule is alive for the indices except the
exponent of $\theta$.

For the purpose of determining the compensating
factor $\theta^A(x)$ 0-form,
we refer to eq.(\ref{current}) in which
$G_M^A({\bf a})_{\mu}$ is composed of
the gauge transformation $\delta_{/{\bf \tau}}{\bf a}^A$
1-form as in eq.(\ref{gg}), and surely
correspond to $\theta^A(x)$ 0-form.
Therefore, $\theta^A(x)$ 0-form must be constructed through the
reduction procedure of $\delta{\bf a}^A$ 1-form.
That is, we need some map $X$:
\be
X:\delta{\bf a}^A \mbox{...1-form}\quad\mapsto\quad\theta^A \mbox{...0-form}.
\ee
The map $X$ can indeed be chosen as
\be
X\equiv{\cal D}^{\dagger},
\ee
where ${\cal D}^{\dagger}$ is the adjoint exterior derivative operator:
\be
{\cal D}^{\dagger}:
\Omega^r(\mbox{M}^2)\rightarrow\Omega^{r-1}(\mbox{M}^2),
\ee
with ${\cal D}^{\dagger}= \ast
{\cal D}\ast$ on the two dimensional Lorentzian manifold
without boundary.
We then obtain $\theta^A(x)$ 0-form as follows:
\be
\theta^A\equiv{\cal D}^{\dagger}\delta{\bf a}^A.
\ee
Accordingly eq.(\ref{add}) is reduced to
\be
\partial_{\beta}(\theta^A R^{A\beta\alpha})=
(\partial_{\beta}\theta^A-(-)^{|B||A|}f_{BA}^{\quad C}
\theta^C{\bf a}_{\beta}^B)R^{A\beta\alpha}
+\theta^A{\cal D}_{\beta}R^{A\beta\alpha},
\label{addpa}\ee
where
\be
{\cal D}_{\beta}R^{A\beta\alpha}=\partial_{\beta}R^{A\beta\alpha}+
(-)^{|B||C|}f_{BC}^{\quad A}{\bf a}_{\beta}^B R^{C\beta\alpha}.
\ee
We next add eq.(\ref{addpa})
to $J_M^0$ which leads to
\be
J_M^0=\sum_{A(\mbox{\scriptsize on $\theta$})}^{all}[\Pi_{
\mu=1}^{A(\mbox{\scriptsize on $\theta$})} R^A
+\theta^A{\cal D}_{\mu=1}R^A],
\label{jj}\ee
where
\be
\Pi_{\mu=1}^{A(\mbox{\scriptsize on $\theta$})} =
(-)^{|B||A|}I_{AB}G_M^B({\bf a})_1+
\partial_1\theta^A-
\sum_C^{all}(-)^{|B||A|}f_{BA}^{\quad C}\theta^C{\bf a}_1^B,
\ee
and $\sum_{A(\mbox{\scriptsize on $\theta$})}^{all}$
denotes a summation of the
indices appearing in the exponent of $\theta$.
That is, we must sum up the indices $A$ in eq.(\ref{jj})
except for the indices appearing in the exponent of the
statistical factor $(-)$.
The index $A$ of $I_{AB}$ and $f_{BA}^C$ in
$\Pi_{\mu=1}^{A(\mbox{\scriptsize on $\theta$})}$
obeys the conventional summation rule.

Our principal task is now to make the topological twist
on the current $J_M^0$ (\ref{jj})
and set it upon the configuration $\Gamma_s$
as a result of the weak coupling limit.
Making the topological twist on $J_M^0$ leads us to
the modified current $J_{\tilde M}^0$
where $\tilde{M} = M+2iV$.
The zero field strengths of the current $J_M^0$ are
removed through replacement of $A$ by $A^{\star}$ (\ref{info}),
which means that  the configuration is reduced to $\Gamma$.
We then obtain the informations for
the limiting condition $J_{\tilde{M}}^0=0$ as follows:
\be
R^{A^{\star}}=0,\qquad\theta^{A^{\star}}=0.
\label{condition}\ee
Clearly, this solution (\ref{condition}) is not unique
in the mathematical view point,
but seems natural because of the independence of the
specific space-time coordinate index: $\mu=1$.

The informations $R^{A^{\star}}=0$ and
$\theta^{A^{\star}}=0$ obtained above play the
roles of the constraints for the configuration ${\Gamma}_s$,
which eventually lead to some moduli space.
Number of the equivariant constraint
$R^{A^{\star}}=0$ is equal to that of the
fermionic connections with ghost number
$\psi_{\mu}(-1)$,\enskip $\psi_{\mu -}^{\quad +}(1)$,\enskip
$\phi_{\mu}(1)$ and $\phi_{\mu -}^{\quad +}(-1)$.
Therefore, $R^{A^{\star}}=0$ can be regarded as the fixing condition
of the so-called topological symmetry whose degrees of freedom
is equal to number of these fermionic connections,
i.e. the so-called topological ghosts.

Let us next explain the physical meaning of
the condition $\theta^{A^{\star}}=0$.
The tangent of the connection space ${\cal A}$
can be decomposed\cite{o.babelon&c.m.viallet1} as follows:
\be
T_{\bf a}{\cal A}=Im{\cal D}\oplus Ker{\cal D^{\dagger}},
\ee
where $Im{\cal D}$ 1-form is the tangent in the gauge
direction, while $Ker{\cal D}^{\dagger}$ 1-form means the
component orthogonal to the gauge orbit and $0=\theta^{A^{\star}}=
{\cal D}^{\dagger}\delta{\bf a}^{A^{\star}}$ is
natural gauge condition in which $\delta{\bf a}^{A^{\star}}$ is
infinitesimal variation of the connection ${\bf a}^{A^{\star}}$.
The constraint $\theta^{A^{\star}}$, the number of which is equal
to that of the generators of the gauge symmetry,
is then regarded as the gauge fixing condition.

We can therefore claim that
all the constraints $R^{A^{\star}}=0$,\enskip$\theta^{A^{\star}}=0$ which
originate from $J_{\tilde{M}}^0=0$ indeed lead to a moduli space
of flat connections:
\be
{\cal M}_{flat}=\{R^{A^{\star}}=0\}/{\cal G^{\star}}.
\label{modu}\ee
The moduli space (\ref{modu}) is really associated with the topological
algebra (\ref{topoal}).
The BRST gauge fixing is necessary,
by way of parenthesis,
for the detailed investigation of observables,
correlation functions and
their geometrical meaning in TFT\cite{h.kanno1}\cite{l.baulieu&i.singer1}.
Incidentally let us describe another representation
for the conditions (\ref{condition}).
If the infinitesimal variation of the connection $\delta{\bf a}^{A^{\star}}$
are on $\Gamma_s$, the variations of $R^{A^{\star}}$
under $\delta{\bf a}^{A^{\star}}$ must also vanish.
Linearized representation\cite{m.f.atiyha&r.bott1}\cite{n.j.hitchin1}
of the flat connection equations yields
\be
\begin{array}{l}
0=\ast\delta R=\ast{\cal D}\delta {\bf a}=
\ast{\cal D}\ast\ast\delta{\bf a}={\cal D}^{\dagger}\ast\delta{\bf a},\\
0={\cal D}^{\dagger}\delta{\bf a}.
\end{array}
\ee
If $\delta{\bf a}$ is the arbitrary variation on ${\cal M}_{flat}$,
its Hodge dual $\ast\delta{\bf a}$, which is
still 1-form only in the two-dimension, is also on ${\cal M}_{flat}$.

Let us next refer to the general coordinate transformations.
In the ordinary N=2 conformal supergravity, all
curvatures must vanish in full consonance with the general coordinate
transformations as gauge symmetry generated by the conformal
super group\cite{supergravityconstraint}.
As a consequence, there exists no kinetic term, i.e.
no dynamics of connection fields in the ordinary theory.
In the present case, on the contrary,
zero curvatures play the roles of the conditions
which lead to the configuration of the fields.
Accordingly, the general coordinate transformation
$\delta_{gc}(\xi)$ is
induced by these conditions.
$\delta_{gc}(\xi)$ expressed as
\be
\delta_{gc}(\xi){\bf a}_{\mu}^A=\sum_B\delta_B(\xi^{\nu}{\bf a}_{\nu}^B)
{\bf a}_{\mu}^A+\xi^{\nu}R_{\nu\mu}^A.
\label{geneco}\ee
The topological twist on eq.(\ref{geneco}) induces
the replacement $A\enskip \rightarrow \enskip A^\star$,
so that the zero field strengths associated with
$A-A^\star$ are removed. Moreover, under the weak coupling
limit, the resultant configuration is given by (\ref{condition}).
Therefore, the transformation law (\ref{geneco}) on ${\cal M}_{flat}$
is described as follows:
\be
\delta_{gc}(\xi){\bf a}_{\mu}^{A^{\star}}=
\sum_{B^{\star}}\delta_{B^{\star}}
(\xi^{\nu}{\bf a}_{\nu}^{B^{\star}}){\bf a}_{\mu}^{A^{\star}},
\label{geneco-on}
\ee
In eqs.(\ref{geneco-on})$, \delta_{gc}(\xi)$ have
been fixed in accordance with
the thoroughly fixed gauge symmetry (\ref{modu}).
It is then possible to argue that
the configuration ${\cal M}_{flat}$ is a quotient
not only in the sense of the gauge symmetry,
but also in the sense of the diffeomorphism:
\be
\sim / {\cal G}^{\star}\quad \supset \quad \sim / \mbox{Diff}_0.
\ee \\

\section{\rm Geometrical Meaning of Fermionic Operators}
\hed In the last section, the moduli space ${\cal M}_{flat}$ (\ref{modu})
has been derived formally.
It is possible to obtain more informations on ${\cal M}_{flat}$
by studying a geometrical meaning of the fermionic operators of
the algebra (\ref{topoal}).
In TFT, the operator $\delta_f$ of the BRST-like fermionic
symmetry corresponds to the exterior derivative operator $d$
on a moduli space where the ghost number corresponds to
the form degree. In the topological
Yang-Mills theory on 4-manifolds\cite{e.witten1}\cite{e.witten3},
for instance, the cotangent vector, i.e. 1-form on the Yang-Mills
instanton moduli space, is described as
\be
\delta_f {\bf a}=\psi,
\ee
where ${\bf a}$ is a generic point in the moduli space and
$\psi$ is a topological ghost.
In the present case,
it is natural to regard the fermionic
operators $Q$, $S$, $Q_c$ $S_c$ as $\delta_f$,
because there exists the ghost number
which has nothing to do with the gauge symmetry
and moreover the fermionic operators generate the transformations
with the ghost number,
that is, they are the ghost number-carriers.
Whatever the gauge orbit may be collapsed under the zero
coupling limit, the four fermionic operators can
still remain as BRST-like operators on ${\cal M}_{flat}$.
The operation on ${\cal M}_{flat}$ must be
read off from the transformation rule (\ref{gaugeal}).
The existence of such operators on ${\cal M}_{flat}$
leads us to consistent and interesting results.
A space which we can regard as a moduli space
will be equiped with some analytical, or in other word,
differentiable structure, in general.
It is well-known, for instance, that the moduli space of
4D (anti-) instantons is locally homeomorphic
to a differentiable manifold under the appropriate conditions
and the complex structure of the (anti-) instanton
moduli space corresponds to that of the basemanifold
\cite{m.itoh1}\cite{a.galperin&o.ogievetsky1}.
It is a matter of course that the topological invariants of TFT,
which are originated from the Donaldoson theory,
must be integrals on certain analytical
support of the moduli space.

We now suppose that the moduli
space ${\cal M}_{flat}$ (\ref{modu})
has such an analytical support.
This assumption is appropriate
becouse the general discussion of flat connections
shows that the moduli space of
flat connections is regarded as a manifold.
In the present case,
it is possible to decompose ${\cal M}_{flat}$ locally
into fermionic sub-space ${\cal M}^f$ and bosonic
sub-space ${\cal M}^b$ as follows:
\be
{\cal M}_{flat} \cong {\cal M}^f \otimes {\cal M}^b.
\ee
Accordingly, ${\cal M}_{flat}$ is regarded as a fiber bundle
over the base space ${\cal M}^b$,
which is described by using the following fibration:
\be
\begin{array}{ccc}
F &
\longrightarrow &
\begin{array}[t]{c}
     {\cal M}_{flat}\\
     \mapdown{\pi} \\
     {\cal M}^b
\end{array}
\end{array}
\label{fibration}\ee
$F$ denotes a fiber parametrized by the ghost number.
That is, ${\cal M}_{flat}$ is regarded as the Whitney sum bundle
composed of the vector bundles;
$E_k$ with the ghost number $k$$(=-1,0,1)$:
\be
{\cal M}_{flat}=E_{-1} \oplus E_0 \oplus E_1.
\ee

Let us discuss a geometrical meaning of the fermionic
operators in the present theory.
In sect.3, the three TFT's
Lagrangians have been set up in (\ref{def-of-lagrangians}).
While, under the weak coupling limit,
we have induced the reduction to the moduli space ${\cal M}_{flat}$
by considering one of the three Lagrangians; ${\cal L}_{QS}$,
all Lagrangians are adapted here.
We introduce a total Lagrangian ${\cal L}_{tot}$ as
a linear combination of the three Lagrangians:
\be
{\cal L}_{tot} ={\cal L}_Q + {\cal L}_S + {\cal L}_{QS}.
\label{tot-lag}\ee
Using the algebra (\ref{topoal}),
${\cal L}_{tot}$ is also described as an anti-commutator
of two fermionic operators as follows:
\be
{\cal L}_{tot} = \{ {\cal QS}, {\cal QS}^\dagger \},
\label{another-tot}\ee
where ${\cal QS} = Q + S_c$ and ${\cal QS}^\dagger = Q_c + S$.
As can be seen from the relations (\ref{i}),
${\cal QS}$ increases the ghost number by one unit,
while ${\cal QS}^\dagger$ decreases it by the same quantity.
Moreover, both ${\cal QS}$ and ${\cal QS}^\dagger$
are nilpotent.
Under the weak coupling limit,
${\cal QS}$ and ${\cal QS}^\dagger$ can be regarded
as operators on the fiber bundle ${\cal M}_{flat}$.
To be precise, ${\cal QS}({\cal QS}^{\dagger})$
operates on the vector bundles $E_k (k=-1,0,1)$.
The operation sequence of ${\cal QS}$ is
\be
\begin{array}{ccccccccc}
0 &
\mapright{i} &
E_{-1} &
\mapright{{\cal QS}_{-1}} &
E_0 &
\mapright{{\cal QS}_0} &
E_1 &
\mapright{{\cal QS}_1} &
0,
\end{array}
\label{sequence}\ee
where $E_{-1} = \{ \psi, \phi^z \} $, $E_0 = \{ \omega, b, e^z, f^z \}$,
$E_1 = \{ \psi^z, \phi \} $,
${\cal QS}_k$ ($k= -1, 0, 1$) $=$ ${\cal QS}$
and $i$ denotes inclusion.
Therefore, we can regard the above sequence (\ref{sequence})
as an elliptic complex, and ${\cal QS}$ as a Fredholm operator
with ${\cal QS}^\dagger$ its adjoint.
Accordingly, ${\cal L}_{tot}$ corresponds to
the Laplacian operator.

It is possible to derive the index of the above elliptic complex
as an additional result:
\begin{eqnarray}
\mbox{ind(${\cal QS}$)} & = & \sum_{k=-1}^1 (-1)^k \mbox{Harm}^k
({\cal M}_{flat}, {\cal QS}) \nonumber \\
& = & 0,
\label{index}\end{eqnarray}
where $\mbox{Harm}^k ({\cal M}_{flat},\enskip {\cal QS})$
denotes the Kernel of $ \{ {\cal QS}_k, {\cal QS}_k^\dagger \}.$
The above discussions may be associated with the Morse theory
applied to the supersymmetric NL-$\sigma$ model\cite{e.witten6}
by E. Witten.
In Ref.\cite{e.witten6}, the Hamiltonian is composed of
the supercharges which correspond to
the coboundary operator and its adjoint of the de Rham theory
and play a crucial role for studying the supersymmetry breaking.
Clearly, ${\cal L}_{tot}$ does not correspond to
the Hamiltonian $H$ in contrast with the theory in Ref.\cite{e.witten6},
because the ghost number
runs from $-1$ to $+1$ and then
${\cal QS}$ can not be identified with the exterior derivative
operator of the de Rham theory.
Therefore, the index and the harmonic forms in (\ref{index})
can not be also identified with the Euler number and
the Betti numbers, respectively.

The vanishing Lagrangian condition
${\cal L}_{tot} = 0$ means that the eigenvalues of the
Laplacian operator must be zero.
The fact seems to show that ${\cal M}_{flat}$
is corresponding to the Kernels;
$\mbox{Harm}^k$(${\cal M}_{flat}$, ${\cal QS}$).
To see the contents of the Kernels,
we describe the operator sequence (\ref{sequence})
in terms of the components:\\
\be
\begin{array}{lllll} \vth \vth
\phi^z &
\displaystyle{\mathop{\rightleftharpoons}_{S}^{Q}} &
f^z &
\mapright{S_c} &
\phi \\ \vth \vth
\psi &
\mapleft{S} &
\omega &
\mapright{Q} &
\phi \\ \vth \vth
\psi &
\mapleft{Q_c} &
e^z &
\displaystyle{\mathop{\rightleftharpoons}_{S}^{Q}} &
\psi^z \\ \vth \vth
{}~ & ~ & b & ~ & ~
\end{array}
\label{component-sequence}\ee
As can be seen from the sequences (\ref{component-sequence}),
four components $e^z$, $f^z$, $\psi^z$ and $\phi^z$
are not suitable for the Kernel of the Laplacian.
In the above context,
we have indicated that ${\cal L}_{tot}$
is regarded as the Laplacian under the weak coupling limit
and the vanishing Lagrangian condition leads us to
the Kernels of the Laplacian.
It is natural to expect that the Kernels correspond
to ${\cal M}_{flat}$.
Consequently, we claim that
the four components must not be
in ${\cal M}_{flat}$.

This statement is also supported by the following consideration.
First of all, there are two closed loops in
the sequences (\ref{component-sequence}), which
are composed of two sets ($e^z$, $\psi^z$), ($f^z$, $\phi^z$):
\be
\begin{array}{ccccccc}
\phi^z &
\displaystyle{\mathop{\rightleftharpoons}_{S}^{Q}} &
f^z &
\quad , \quad &
e^z &
\displaystyle{\mathop{\rightleftharpoons}_{S}^{Q}} &
\psi^z.
\end{array}\ee
The two components in each set circulate on its own loop
by operation of $Q$ and $S$.
Any other componets can not reach the two loops by any operations.
Moreover, the $S$-operation after the $Q$-operation
(or $Q$ after $S$) leads the component to itself and
such a double operation of $S \ast Q$ (or $Q \ast S$)
induces the transformation
in the direction of the gauge orbit,
because of the relation $[Q,S] = -4i{\tilde M}$.
Therefore, the two closed loops are in the gauge orbit
and are expected to be collapsed together with
the gauge orbit on ${\cal M}_{flat}$.
The fact shows that
$e^z$, $f^z$, $\psi^z$ and $\phi^z$
are not in the muduli space ${\cal M}_{flat}$.

More detailed informations of ${\cal M}_{flat}$ are obtained,
through studying the geometrical meaning
of the fermionic operators.
It is clarified that
the choice of the left-chiral part of the algebra (\ref{topoal})
is rather meaningless for the purpose of the
derivation of the moduli space and
the intersection of the left- and right-chiral part is only effective.
Therefore, ${\cal M}_{flat}$
is reduced to ${\cal M}_0$:
\be
{\cal M}_0 = \{ R^{A^{\star \star}}=0 \} / {\cal G}^{A^{\star \star}},
\ee
where $A^{\star \star} = {\tilde M}, {\tilde D}, Q, S$.
We then claim that
the moduli space intrinsic to the topologically twisted
\osp is really ${\cal M}_0$.\\

\befl \underline{ 4-1 \ Observable} \efl

Let us mention the triviality of the path-integral.
The triviality depends on the existence and
the characterestics of observables.
{}From the view point of the ordinary quantum gauge theory,
the condition for observable is the gauge invariance.
In TFT, another condition is required, i.e.
metric independence
as well as gauge invariance.
The above two conditions would not be enough to yield non-trivial
correlation functions.
The fermionic contribution in the path-integral measure
must not be disregarded.
If fields are assigned by ghost number,
the path-integral measure may have ghost number anomaly.
Therefore, to absorb the ghost number of the measure,
observables must have the same ghost number as the measure.

In the present case, while we see that the measure is free from
the ghost number anomaly,
the measure on the moduli space ${\cal M}_0$ includes the
fermionic contributions $d\psi \, d\phi$
and the difficulty of Grassman number integration is retained.
If the integrand does not include the coupling $\psi \, \phi$,
the integration on ${\cal M}_0$ is zero.
After all, the triviality of the path-integral
depends on the existence of observables which
are gauge invariant, metric independent and
including the coupling $\psi \phi$,
rigorously speaking, gauge invariance and metric independence
modulo ${\cal Q}$-exact.

Now let us find the observable which satisfies
the above three conditions.
We note that the TFT's
Lagrangians (\ref{def-of-lagrangians}) are
invariant under the gauge symmetry generated
by the algebra (\ref{topoal})
composed of $A^\star$ (\ref{info})
at the quantum level.
First of all, we must pay attention to the
Yang-Mills action associated with the algebra $A^\star$:
\be
{\cal O}_{\mbox{\scriptsize YM}_2}
=\int_{\mbox{\scriptsize M}^2} I_{AB}R^A\wedge \ast R^B,
\label{yangmills}\ee
where $A$($B$) is the label of the algebra $A^\star$.
The above $I_{AB}$ denotes the Cartan-Killing matrix:
\be
I_{AB}=-(-)^{|A|+|C|(|A|+|B|+1)}f_{AD}^{\quad C}f_{BC}^{\quad D}.
\label{cartankilling}\ee

The Yang-Mills action in two dimensions:
\be
{\cal I}_{\mbox{\scriptsize YM}_2}=
\int_{\mbox{\scriptsize M}^2}
d\mu g^{ai}g^{bj}\mbox{Tr}(F_{ab}F_{ij})
\ee
do not depend on
a metric $g$ in general\cite{e.witten7}.
The metric dependence is up to the measure $d\mu$
on the two dimensions $\mbox{M}^2$.
A curvature two form $R$ can be written as
$R=\varepsilon_v \hat{R}$ where $\varepsilon_v$
denotes an two form determined by a metric
and $\hat{R}$ is an algebra-valued zero form.
Using the zero form $\hat{R}$, the Yang-Mills action ${\cal I}_{\mbox{
\scriptsize YM}_2}$ can be described as
\be
{\cal I}_{\mbox{\scriptsize YM}_2}=\int_{\mbox{\scriptsize M}^2}
d\mu I_{AB} \hat{R}^A \hat{R}^B.
\label{metricindep}\ee
Therefore, we can regard ${\cal I}_{\mbox{
\scriptsize YM}_2}$ as metric independent
if $d\mu$ and $\hat{R}$ do not contain a metric.
It is a matter of course that the requirement of
coordinate transformation invariance of $d\mu$ would induce
metric dependence of $d\mu$, however.

Now our concentration on the Yang-Mills action
${\cal O}_{\mbox{\scriptsize YM}_2}$ (\ref{yangmills})
will be recoverd.
We then evaluate the Cartan-Killing matrix $I_{AB}$.
Number of the generators of $A^\star$ is $8$ and
number of the independent elements
of the Cartan-Killing matrix $I_{AB}$ is then $28$.
Using the commutation relations (\ref{topoal})
of the algebra $A^\star$ and the definition (\ref{cartankilling})
of $I_{AB}$, we can see that
the Cartan-Killing matrix $I_{AB}$ is zero matrix.
Therefore, the Yang-Mills action becomes zero and
we must find another candidate for observable.

Let us consider next candidate.
The metric independent integral (\ref{metricindep})
will be still useful.
If $I_{AB} \equiv 1$, the two dimensional integration
\be
{\cal O}_{\hat{R}^A \hat{R}^B}=\int_{\mbox{\scriptsize M}^2}
d\mu \hat{R}^A \hat{R}^B
\ee
will not be invariant under the gauge symmetry,
in contrast to the Yang-Mills action.
Notwithstanding, we expect that some
${\cal O}_{\hat{R}^A \hat{R}^B}$ will be
gauge invariant modulo $\delta_{\cal Q}$-exact.
First of all, we must pay attention to the fact that
all $\hat{R}^A$ can be represented as ${\cal Q}$-exact form:
\be
\hat{R}^A\sim \delta_{\cal Q} \hat{R}^B,
\label{exact}\ee
where the {\sc rhs} of eq.(\ref{exact}) is not necessarily unique
due to the algebra $A^\star$ (\ref{topoal}).
For instance, $\hat{R}^{\tilde M}\sim
\delta_{Q_c} \hat{R}^{Q_c}\sim
\delta_{S_c}\hat{R}^{S_c}$.
The ${\cal Q}$-exact form (\ref{exact}) leads to
the fact that all quadratic forms
$\hat{R}^A \hat{R}^B$ except for $\hat{R}^P \hat{R}^K$
and $\hat{R}^{Q_c} \hat{R}^{S_c}$ can be
also described as the ${\cal Q}$-exact form:
\be
\hat{R}^A \hat{R}^B \sim
\delta_{\cal Q} \hat{R}^C \hat{R}^D.
\label{quadraexact}\ee
The two quadratic forms $\hat{R}^P \hat{R}^K$
and $\hat{R}^{Q_c} \hat{R}^{S_c}$ are
related as follows:
\be
\hat{R}^P \hat{R}^K \sim \delta_{{\cal Q}_1} (\hat{R}^{Q_c}
\delta_{{\cal Q}_2} \hat{R}^{S_c}) +
a \hat{R}^{Q_c} \hat{R}^{S_c},
\label{pkqs}\ee
where $a$ is some constant determined by the algebra $A^\star$.
${\cal Q}_1$ and ${\cal Q}_2$ are
i) $Q$ and $S$, respectively, or ii) $S$ and $Q$, respectively.
Then we see that ${\cal O}_{\hat{R}^P \hat{R}^K}\sim
{\cal O}_{Q_c S_c}$
in the path-integral.
While ${\cal O}_{\hat{R}^P \hat{R}^K}$ (or equivalently
${\cal O}_{\hat{R}^{Q_c} \hat{R}^{S_c}}$)
is non-trivial and metric independent,
unfortunately ${\cal O}_{\hat{R}^P \hat{R}^K}$
is not gauge invariant
(mudulo $\delta_{\cal Q}$-exact):
\be
\delta_g \hat{R}^P \hat{R}^K \sim
b \hat{R}^P \hat{R}^K
+ \delta_{\cal Q}\mbox{-exact},
\label{gonext}\ee
where $b$ is some constant determined by
the algebra $A^\star$.
Therefore, the second candidate
of the form ${\cal O}_{\hat{R}^A \hat{R}^B}$ can
not be accepted as observable.

Our final trial to obtain the non-trivial observable
will be as follows.
The first candidate ${\cal O}_{\mbox{\scriptsize YM}_2}$
occur easily to us, which results to be zero;
${\cal O}_{\mbox{\scriptsize YM}_2}=0$.
The second candidate is the modification of the first candidate.
On the contrary, the final candidate which we will show
is somewhat heuristic.
Let us begin with introducing the following two forms:
\be
\varrho_{\alpha }=\varepsilon_v
\hat{\varrho}_{\alpha }, \enskip (\alpha=0,1).
\label{varrho}\ee
$\varepsilon_v$ denotes the above mentioned
two form determined by a metric and
$\hat{\varrho}_{\alpha}$ is zero form defined by
\be
\hat{\varrho}_{\alpha}=\psi_{\alpha} \phi_{\alpha}.
\label{hat}\ee
The index $\alpha$ can be regarded as mere labels
because the coordinate transformations are not considered.
Therefore, we present the final
candidate as in the following:
\be
{\cal O}_{\varrho}=\int_{\mbox{\scriptsize M}^2}
d\mu \; \varrho_0 \varrho_1.
\label{final}\ee
The behavior of $\varrho_{\alpha}$ under
the gauge transformation $\delta_g$ (\ref{gaugeal})
is written in the form:
\begin{eqnarray}
\delta_g \varrho_{\alpha} & = &
                 (\partial_{\alpha} \epsilon ) \phi_{\alpha}+
                 \psi_{\alpha} \partial_{\alpha} \kappa
                                                                      \nonumber
\\
& = & \acute{\epsilon} \phi_{\alpha} - \acute{\kappa} \psi_{\alpha}
\nonumber \\
& = & \delta_{QS} \; \omega_{\alpha},
\label{ozuha}\end{eqnarray}
where $\acute{\epsilon}=\partial_{\alpha} \epsilon$
and $\acute{\kappa}=\partial_{\alpha} \kappa$.
Therefore, we see that
\begin{eqnarray}
\delta_g {\cal O}_{\varrho} & = &
  \int_{\mbox{\scriptsize M}^2} d\mu \;
   \delta_g ( \varrho_0 \varrho_1 ) \nonumber \\
& = & \int_{\mbox{\scriptsize M}^2} d\mu
          \{ ( \delta_g \; \varrho_0 ) \varrho_1
              + \varrho_0 ( \delta_g \; \varrho_1 ) \} \nonumber \\
& = & \int_{\mbox{\scriptsize M}^2} d\mu
         \{ ( \delta_{QS} \; \omega_0 ) \varrho_1
           + \varrho_0 ( \delta_{QS} \; \omega_1 ) \} \nonumber \\
& = & \int_{\mbox{\scriptsize M}^2} d\mu \;
\delta_{QS} \; ( \omega_0 \varrho_1 + \varrho_0 \omega_1
                      - \omega_0 \omega_1) \nonumber \\
& = & \delta_{QS} \int_{\mbox{\scriptsize M}^2} d\mu
           ( \omega_0 \varrho_1 + \varrho_0 \omega_1
             - \omega_0 \omega_1).
\label{rippa}\end{eqnarray}
The above eq.(\ref{rippa}) shows that
${\cal O}_{\varrho}$ is gauge invariant in the path-integral.
Clearly, ${\cal O}_{\varrho}$ is not dependent on a metric
and includes the coupling
$\psi_0 \, \phi_0 \, \psi_1 \, \phi_1$.
Therefore, we conclude that
${\cal O}_{\varrho}$ is a
non-trivial TFT's observable.
The fact shows that the path-integral is not trivial
and supports the present discussion from beginning to end.\\
\section{\rm Summary and Remarks}
\hed {\it First}: We have investigated the topologically twisted
\osp conformal superalgebra and
derived the moduli space intrinsic to the twisted algebra.
The algebra includes the appropriate TFT's Lagrangians
composed of the fermionic charges
$Q$, $S$, $Q_c$ and $S_c$.
They lead us to the moduli space ${\cal M}_{flat}$
intrinsic to the algebra
under the condition of the weak coupling limit.
As consequence of the investigation of
the geometrical meaning of the fermionic charges,
it is shown that ${\cal M}_{flat}$ is reduced
to ${\cal M}_0$ associated with the intersection of the
left- and right-chiral part of the topological algebra
and is just a moduli space inherent to the algebra.
As an additional result, the index of these fremionic operators
is derived if some proper support in
the moduli space can be defined.
The facts which have been clarifed in the above discussion
show that the topological algebra
has a specific relation
with a moduli problem.
It is claimed that a geometrical feature of the algebra
is one of the interesting characteristics inherent to
the topological twist.
Therefore, we has succeeded in shedding some light upon
the relation between
the topological twist and the moduli problem
through the geometrical aspect of the topological algebra.
{\it Secondly}:
Let us make a remark on
the vanishing Noether current.
Making use of the ambiguity of the Noether current,
we are led to the relation (\ref{jj}) between the vanishing Noether
current $J_{\tilde{M}}^0=0$ and the flat connection
conditions $R^{A^{\star}}=0, \theta^{A^{\star}}=0$.
In the conventional QFT, e.g. the quantum electro dynamics,
this ambiguity mentioned above
plays an important role in association with
avoiding one mass-less state to obtain
a well-defined conserved charge while its physical role is
not clarified in the case of the classical correspondent.
In the present TFT, on the contrary,
the classical theory has been obtained as
the limiting case of the path-integral,
and consequently the ambiguity argued above leads to
the corresponding moduli problem.

\vspace{1.7cm}

{\it Acknowlegment}:
The author would like to thank Professor H.Fujisaki
for encouragement,
enlightening discussions and careful reading of the manuscript.
He is grateful to Dr.T.Suzuki(Yukawa Inst.)
for helpful comments
at the early stage of this work.
\\


\end{document}